\documentclass[prb,preprint]{revtex4-1}

\usepackage{amsmath}  
\usepackage{amsfonts} 
\usepackage{graphicx} 
\usepackage[colorlinks,linkcolor=blue,anchorcolor=blue,citecolor=blue]{hyperref}
\newcommand{\upcite}[1]{\textsuperscript{\textsuperscript{\cite{#1}}}}
\usepackage[svgnames]{xcolor}
\usepackage[framed]{ntheorem}
\usepackage{framed}
\usepackage{tikz}

\begin{document}
    \theoremclass{Theorem}
    \theoremstyle{break}


\theorembodyfont{}

\theoremseparator{\hspace{1em}}

\theoremnumbering{arabic}
\newtheorem{Theorem}{Theorem}
\newtheorem{Definition}{Definition}
\newtheorem{Cor}{Corollary}[Theorem]
    \tikzstyle{thmbox} = [rectangle, rounded corners, draw=black,
      fill=white
      , inner sep=10pt 
      ]
    \newcommand\thmbox[1]{%
      \noindent\begin{tikzpicture}%
      \node [thmbox] (box){%
        \begin{minipage}{.8\columnwidth}%
          #1%
        \end{minipage}%
      };%
      \end{tikzpicture}}

    \let\theoremframecommand\thmbox
    \newshadedtheorem{ImportantTheorem}[Theorem]{Theorem}
    \newshadedtheorem{ImportantDefinition}[Definition]{Definition}
    \newshadedtheorem{ImportantCor}[Cor]{Corollary}


\title{The Equilibrium Statistical Model of Economic Systems\\ using Concepts and Theorems of Statistical Physics}
\author{Zhiwu Zheng}
\affiliation{Department of Physics, Nanjing University, Jiangsu, China 210093}





\begin{abstract}
Economic systems are similar with physic systems for their large number of individuals and the exist of equilibrium. In this paper, we present a model applying the equilibrium statistical model in economic systems. Consistent with statistical physics, we define a series of concepts, such as economic temperature, economic pressure, economic potential, wealth and population. Moreover, we suggest that these parameters show pretty close relationship with the concepts in economy. This paper presents related concepts in the equilibrium economic model and constructs significant theorems and corollaries, which is derived from the priori possibility postulate, getting theorems including the equilibrium theorem  between open systems, the distribution theorem of wealth and population along with related corollaries. More importantly, we demonstrate a method constructing optimal density of states, optimizing a macroscopic parameter depending on need to get the distribution of density of states utilizing the variation method, which is significant for the development of a society. In addition, we calculate a simple economic system as an example, indicating that the system occupied mostly by the middle class could develop stably and soundly, explaining the reason for resulting distributions of macroscopic parameters.
\end{abstract}

\maketitle 

\section{Introduction} 
In statistical physics\upcite{1}, the statistical model originated from physics does not restrict exactly the type of the system. It means that this model can not only applied on the real physical systems such as moving particles, localized lattice, polarized molecules but also systems consisting of more metaphysical particles (can also be called subsystems or quasi-particles). The \textit{economic system} is one of these.

Combined economics with physics, a lot of work have been done\upcite{2}. These work mainly concentrate on the changes in price (Power laws in 1897\upcite{3}\upcite{4} and Levy Distribution\upcite{5} in early 1960s), the equilibrium in economic systems making attempts to explain the economic growth (DGSE Model in 1990s\upcite{6}), analysis on price and stock markets using Brownian  motion and random walks (first developed by Louis Bachelier\upcite{7} in 1900), and the use of Ising Model to account for social influence in individual decisions (Phan et al\upcite{8} (2004)), the decision-making using Boltzmann distribution (Luce and Suppes \upcite{9} (1965)). However, there is few thesis putting their attention on more fundamental concepts describing the properties and its interaction with outer systems, such as temperature, pressure, volume, economic potential and the partition function.

Economic system, as the same as the conventional concept, is an assemble of economic subsystems with interaction. The subsystem can not only be refereed to a human being, but also an economic man such as a company and a shop. The economic system such as a nation usually consists of a large number of subsystems, leading to small fluctuations. Moreover, in the research of economics, it is recognized the economic system can achieve the equilibrium (such as the equilibrium of commodity market, money market and labor market). So it can be recognized that the equilibrium statistical model of physics can be applied to economic systems to research on the properties of one equilibrium economic system and what is the conversion like between two such systems. In this paper, we present a model of economic system applying the physics statistical model, giving basic concepts and theorems along with building basic frames of this field. At first, the basic definitions, axioms, and theorems will be given, then with the meaning of each quantity. And we will present some direct conclusions of this model by giving one example using it.

\section{Model}\label{model}
\subsection{Definitions of Some Quantities}
\begin{table}[htb]
\begin{center}
\caption{The Names And Symbols of Some Quantities}\label{SymNam}
\begin{tabular}{|c|l|}
\hline
Symbol & Name\\
\hline
T & Economic temperature\\
$\mu$ & Economic potential\\
V & Economic volume\\
U/E & Wealth \\
p & Economic pressure\\
$\Omega$ & Economic microstate number\\
$\mathcal{Z}$ & Economic grand partition function\\
- & Economical individual\\
- & Isolated economic system\\
- & Open economic system\\
\hline
\end{tabular}
\end{center}
\end{table}

These quantities will give their definitions when introduced.And they are usually decorated with an "economic" to distinguish with those of physics.

\begin{ImportantDefinition}[Economic Volume(V)]
{\textit Economic Volume} is used to measure the quantity of resources occupied by an economic system.
\end{ImportantDefinition}
\begin{ImportantDefinition}[Wealth]
\textit{Wealth}, as the same as usually defined, refers to the sum of goods' value of a subsystem. It does not include the resources not developed. When the economic system reaches equilibrium, the wealth does not vary with time.
\end{ImportantDefinition}
\begin{ImportantDefinition}[Economic Microstate Number~$\Omega$£©]
\textit{Economic microstate number} is the number of all possible microstates varied with the number of economic individuals and individual wealth.
\end{ImportantDefinition}
\begin{ImportantDefinition}[Isolated Economic System]
The \textit{isolated economic system} is the system which does not transfer wealth or individual to the outside world.
\end{ImportantDefinition}
\begin{ImportantDefinition}[Open Economic System]
The economic system transferring wealth and individual freely is the \textit{open economic system}.
\end{ImportantDefinition}

\subsection{Basic Postulation and Developed Theorems}
The foundation of this model is the postulation \textit{equal a priori probability postulate}, which makes each microstate occurs in a same possibility in isolated systems. In other words, wealth distributed uniformly in a isolated economic system. However, this uniform wealth distribution is different from what we understand normally. It means that although it is equally possible for each microstate, it will not lead to equal wealth of each economic individual. To clarify this statement, we can introduce the concept \textit{degeneracy}, referring to the number of microstates who have the same quantity of wealth one wealth level can hold. Hence, the equal a priori probability postulate require economic individuals have the same opportunity to develop in each aspect. \\

This postulation leads to the most portable distribution, the distribution holds largest number of microstates, along with the following definitions, theorems and corollaries as usually done in statistical physics\upcite{10}:
\begin{ImportantDefinition}[Economic System's Temperature $T$, Economic potential $\mu$]
In an equilibrium economic system:
\begin{equation}
\frac{1}{T}\triangleq \frac{\partial ln\Omega}{\partial E}
\label{defT}
\end{equation}
\begin{equation}
\frac{1}{\mu} \triangleq \frac{\partial ln\Omega}{\partial N}
\label{defMu}
\end{equation}
\end{ImportantDefinition}
\begin{ImportantTheorem}[The Equilibrium Theorem in Wealth and Population]
When wealth and population (number of economic individuals) of two open economic systems reach its equilibrium, it satisfies that:
\begin{equation}
T_{1}=T_{2}
\label{eq of t}
\end{equation}
\begin{equation}
\mu _{1} =\mu _{2}
\label{eq of mu}
\end{equation}
$T_{1}$, $T_{2}$, $\mu_{1}$, $\mu_{2}$ are economic temperature and economic potential of two systems respectively.
\end{ImportantTheorem}
\begin{ImportantDefinition}[Economic Grand Partition Function]
In an open equilibrium open economic system, with economic temperature $T$, economic potential $\mu$, the economic grand partition function is defined as:
\begin{equation}
\mathcal{Z}\triangleq \sum\limits_{N_{s}}\sum\limits_{E_{s}}e^{-\alpha N_{s}-\beta E_{s}}
\label{defZ}
\end{equation}
\end{ImportantDefinition}
\begin{ImportantTheorem}[The Distribution Theorem of Wealth and Population]
As for an open economic system, the possibility it is with economic temperature $T$, economic potential $\mu$, wealth $E_{s}$, the number of economic individuals $N_{s}$£º
\begin{equation}
P(E_{s},N_{s})=\frac{e^{-\alpha N_{s}-\beta E_{s}}}{\mathcal{Z}}
\label{DistriTh}
\end{equation}
which $\alpha\triangleq \frac{1}{\mu}$, and $\beta\triangleq\frac{1}{T}$.
\end{ImportantTheorem}
\begin{ImportantDefinition}[The Wealth of an Economic System (U)]
\begin{equation}
U\triangleq <E_{s}>
\label{defU}
\end{equation}
\end{ImportantDefinition}
The wealth of an economic system is defined as the average value of all possible states the system can occupy, which is the measurement value from a macroscopic view.
\begin{ImportantCor}[The Wealth of an Equilibrium Economic System]
\begin{equation}
U=-\frac{\partial}{\partial \beta}ln\mathcal{Z}
\label{CorU}
\end{equation}
\end{ImportantCor}
\begin{ImportantDefinition}[The Population of an Economic System ($N$)]
\begin{equation}
N \triangleq <N_{s}>
\label{defn}
\end{equation}
\end{ImportantDefinition}
It is the population of a society (including all economic men) measured from a macroscopic view.
\begin{ImportantCor}[The Population of an Equilibrium Economic System]
\begin{equation}
N=-\frac{\partial}{\partial \alpha}ln\mathcal{Z}
\label{corn}
\end{equation}
\end{ImportantCor}
We are able to believe this: the wealth and population in equilibrium system are optimal.
\section{Meanings of Parameters in an Economic System}
\subsection{The Implication of Economic Pressure}
\begin{ImportantDefinition}[Economic Pressure]
The economic pressure of an equilibrium economic system can be defined as:
\begin{equation}
p\triangleq <(\frac{\partial E_{s}}{\partial V})_{\beta}>
\label{defp}
\end{equation}
\end{ImportantDefinition}

Its mathematical expression could be derived from the partition function.
\begin{equation*}
\begin{aligned}
p&=\sum\limits_{N_{s}}\sum\limits_{E_{s}} \frac{\partial E_{s}}{\partial V} \frac{e^{-\alpha N_{s}-\beta E_{s}}}{\mathcal{Z}}\\
&=-\frac{1}{\beta}\frac{1}{\mathcal{Z}}\frac{\partial}{\partial V}\sum\limits_{N_{s}}\sum\limits_{E_{s}}e^{-\alpha N_{s}-\beta E_{s}}\\
&=-\frac{1}{\beta}\frac{1}{\mathcal{Z}}\frac{\partial}{\partial V}\mathcal{Z}\\
&=-\frac{1}{\beta}\frac{\partial}{\partial V}ln\mathcal{Z}
\end{aligned}
\end{equation*}

It can been seen from Eq.\ref{defp} that the economic pressure refers to the efficiency of utilizing natural resources, the production efficiency in economics.
\begin{ImportantCor}[Economic pressure measures the production efficiency]
The expression of economic pressure is£º
\begin{equation}
p=-\frac{1}{\beta}\frac{\partial}{\partial V}ln\mathcal{Z}
\label{corp}
\end{equation}
Economic pressure measures the production efficiency of an economic system in an macroscopic way.
\end{ImportantCor}

Moreover, through the comparison between two or some economic systems, the relationship in economic pressures could determine whether there would be an invasion in an ideal situation. The ideal situation refers to the invasion behavior would not be prevented by outer world and it is the purpose of all economic systems to maximize their wealth, which actually would lead to the maximum of the sum of wealth.\\

It can be demonstrated as follows: Considering two economic systems A and B, whose economic pressure are $p_{A}$ and $p_{B}$ respectively. When $p_{A}>p_{B}$, A will get more wealth than B does if they get the same quantity of economic volume. If the sum of economic value is fixed, B will be invaded by A. When $p_{A}<p_{B}$, it can be derived by doing the same thing that A will invaded by B. When $p_{A}=p_{B}$, two systems are in dynamic equilibrium, where the two systems do not invade each other from a macro view.\\

\begin{ImportantCor}[The quantity of economic pressure determines the invasion]
In the ideal situation mentioned, considering two equilibrium economic system, the system will be invaded by the other having larger economic pressure.
\end{ImportantCor}

\subsection{The Meaning of Economic Temperature and Economic Potential}
In the section \ref{model}, we build up the definitions of economic temperature and economic potential. When they are different in two systems, the population and the wealth flow as the following corollary:
\begin{ImportantCor}[The Wealth and Population Flow Between Economic Systems]
As for two equilibrium economic systems, the wealth flow from the system with higher temperature to the lower one, and the economic individuals flow from the system with higher economic potential to the lower one.
\end{ImportantCor}
In other words, economic systems with low economic temperature attract wealth and ones with low economic potential attract economic individual. Economic temperature is the quantity measuring wealth attraction and economic potential is the quantity describing the individual attraction.\\

As for an wealth-and-individual changing system, if the speed of temperature and potential changing is much less than the speed the system comes to equilibrium, the changing of the system can been seen as a turn between equilibrium economic systems.\\
\section{The Optimization of the Grand Partition Function and Wealth Distribution}
The above part demonstrates the economic grand partition function (Eq. \ref{defZ}), and derive the wealth of the system (Eq. \ref{CorU}), the population (Eq. \ref{corn}), and economic pressure (Eq. \ref{corp}). However, for a system, we should first know the wealth distribution and then be able to derive the partition function, and finally parameters of the economic system. But through adjustion such as the change of policies, the wealth distribution could be changed freely. Therefore, a question was come up: in order to get the optimal value of one specified parameter, what distribution should be chosen? This takes a crucial effect on the development of society.\\

In this section, the variational method is used to solve this problem.\\
\subsection{The Partition Function on Different Wealth Levels}
Firstly, through the method similar with the Statistical Physics to describe this economic system, assume the economic system is a similar-Boltzmann system, whose each state can occupy as many individuals as possible. \\

For a economic system, there are following relations:
\begin{equation}
\begin{cases}
\sum\limits_{l}a_{l}=N\\
\sum\limits_{l}a_{l}\varepsilon_{l}=E
\end{cases}
\end{equation}
$a_{l}$is the number of individual on the l-th level, $\{a_{l}\}$ is a distribution, called the \textit{occupation number representation}. $\varepsilon_{l}$ is the wealth on l-th level.\\

Therefore, the grand partition function on different wealth levels could be derived:
\begin{equation}
\mathcal{Z}=\prod_{l}\mathcal{Z}_{l}
\end{equation}
where
\begin{equation}
\mathcal{Z}_{l}=\exp[w_{l}e^{-\alpha -\beta \varepsilon_{l}}]
\end{equation}
Hence,
\begin{equation}
\ln\mathcal{Z}=\sum\limits_{l}w_{l}e^{-\alpha -\beta \varepsilon_{l}}
\end{equation}
Applying quasicontinuum method and introducing \textit{density of states} $g(\varepsilon)$, we can get that:
\begin{equation}
\ln\mathcal{Z}= \int_{0}^{\infty}g(\varepsilon)e^{-\alpha -\beta \varepsilon}d\varepsilon
\label{lnz}
\end{equation}
The density of states at 0 and infinity should be zero, that is $g(E)|_{E=0}=0$, $g(E)|_{E=\infty}=0$.\\

Hence, the question mentioned could be expressed. Assume $X$ is the optimal observable, there should be:
\begin{equation}
\delta X=0
\end{equation}
\subsection{The Optimal Density of States Getting Maximum Economic Pressure}
Deduced from Eq. \ref{corp}, \ref{lnz}, the economic pressure can be written as:
\begin{equation}
p=-\frac{1}{\beta}\frac{\partial}{\partial V} \int_{0}^{\infty}g(\varepsilon)e^{-\alpha -\beta \varepsilon}d\varepsilon
\end{equation}

where $V$ and $g$ are functions of $\varepsilon$, that is $V=V(\varepsilon)$, $g=g(\varepsilon)$. Hence,
\begin{align*}
p&=\int_{0}^{\infty}\frac{1}{\beta}V'(\varepsilon)\frac{d}{d\varepsilon}[g(\varepsilon)e^{-\alpha-\beta \varepsilon}]d\varepsilon\\
&=\int_{0}^{\infty}\frac{V'(\varepsilon ) \left(e^{-\alpha -\beta  \varepsilon } g'(\varepsilon )-\beta  g(\varepsilon )\right)}{\beta}d\varepsilon
\end{align*}
Applying Euler's Equation \upcite{11}
, when $\delta p=0$, there would be:
\begin{equation}
\begin{cases}
\frac{e^{-\alpha -\beta  x} [\beta  V'(x) \left(1-e^{\alpha +\beta  x}\right)-V''(x)]}{\beta }=0\\
\frac{e^{-\alpha -\beta  x} [\beta  g'(x) \left(e^{\alpha +\beta  x}+1\right)-g''(x)]}{\beta }=0
\end{cases}
\end{equation}
The solution is:
\begin{equation}
V(\varepsilon)=\frac{c_{1} e^{-\alpha -e^{\alpha +\beta  \varepsilon}}}{\beta }+c_{2}
\label{ve}
\end{equation}
\begin{equation}
g(\varepsilon)=\frac{c_{3} e^{e^{\alpha +\beta  \varepsilon}-\alpha }}{\beta }+c_{4}
\end{equation}
$c_{1}$, $c_{2}$, $c_{3}$, $c_{4}$ are undetermined parameters.
\begin{figure}[htb]
\begin{center}
\includegraphics[width=0.8\columnwidth]{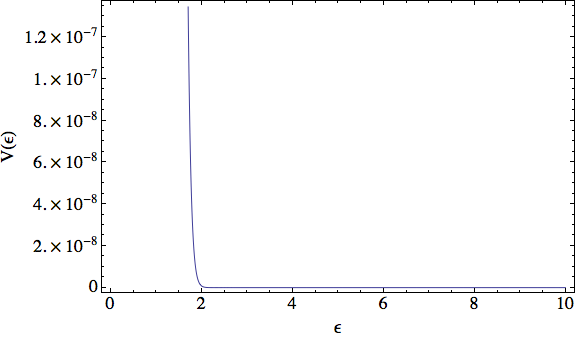}

\caption{The plot of $V(\varepsilon)$ when $\alpha=1$, $\beta=1$, $c_{1}=1$, $c_{2}=0$}
\label{ve}
\end{center}
\end{figure}
\begin{figure}[htb]
\begin{center}
\includegraphics[width=0.8\columnwidth]{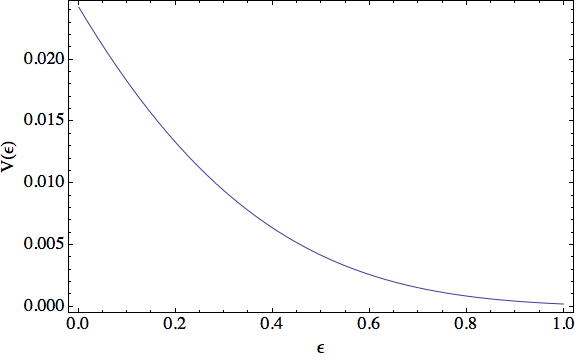}

\caption{The Zoom-in of Fig. \ref{ve} on (0,1)}
\label{veda}
\end{center}
\end{figure}

\begin{figure}[t]
\begin{center}
\includegraphics[width=0.8\columnwidth]{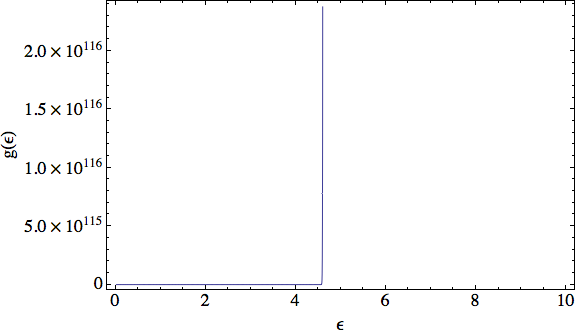}

\caption{The Plot of $g(\varepsilon)$ when $\alpha=1$, $\beta=1$, $c_{3}=1$, $c_{4}=0$}
\label{ge}
\end{center}
\end{figure}

\begin{figure}[htb]
\begin{center}
\includegraphics[width=0.8\columnwidth]{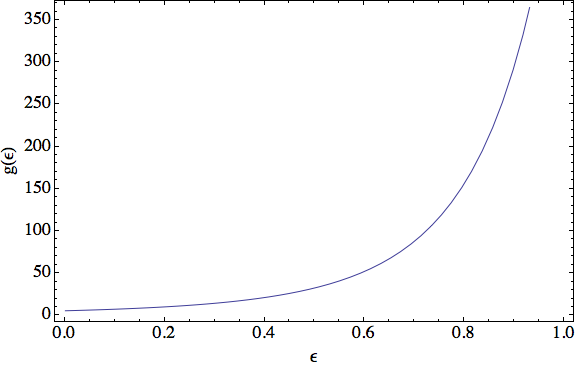}

\caption{The Zoom-in of Fig. \ref{ge} on (0,1)}
\label{geda}
\end{center}
\end{figure}

From Figure \ref{ve} and Figure \ref{ge}, it could be observed that $V(\varepsilon)$ and $g(\varepsilon)$ have opposite trends with increase of $\varepsilon$, where with rasing $\varepsilon$, $V(\varepsilon)$ goes close to 0 with a gradually reduced slope, while $g(\varepsilon)$ increases more rapidly. However, V could not be arbitrarily small, there must be a minimum required economic volume (noted as $b$) as the prerequisite to maintain a economic system. Let $\varepsilon_{0}$ be the related wealth level with $b$, which satisfies $V(\varepsilon_{0})=b$. Therefore, the maximum possible wealth level in this system is $\varepsilon_{0}$, which means that:
\begin{equation}
\sum\limits_{l}w_{l}\rightarrow \int_{0}^{\varepsilon_{0}}g(\varepsilon)d\varepsilon
\end{equation}
\section{Calculation of a Simple Economic System}
\begin{figure}[htb]
\begin{center}
\includegraphics[width=0.8\columnwidth]{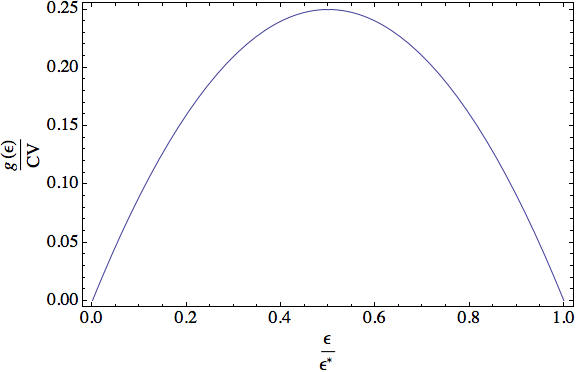}

\caption{The Function of the Density of States}
\label{g}
\end{center}
\end{figure}

We assume that the density of states of a simple economic system is:
\begin{equation}
g(\varepsilon)=-CV\varepsilon(\varepsilon-\varepsilon_{*})
\end{equation}
where $0\leq\varepsilon\leq\varepsilon_{*}$, $C$ is a constant. The plot of this function is shown as Figure \ref{g}.
\subsection{The Grand Partition Function and the Macroscopic Parameters}
\begin{figure}[htb]
\begin{center}
\includegraphics[width=0.8\columnwidth]{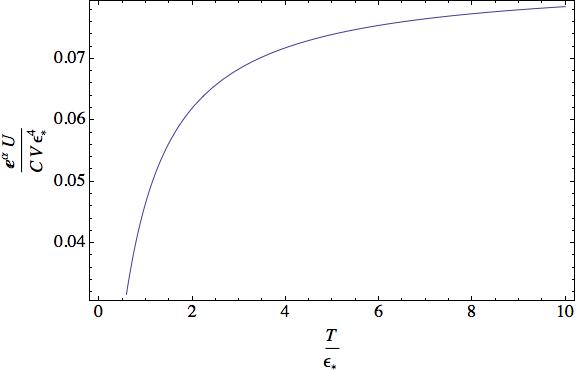}

\caption{The plot of the wealth of the system varied with economic temperature $T$. $T\in (0,10)$}
\label{u}
\end{center}
\end{figure}

\begin{figure}[htb]
\begin{center}
\includegraphics[width=0.8\columnwidth]{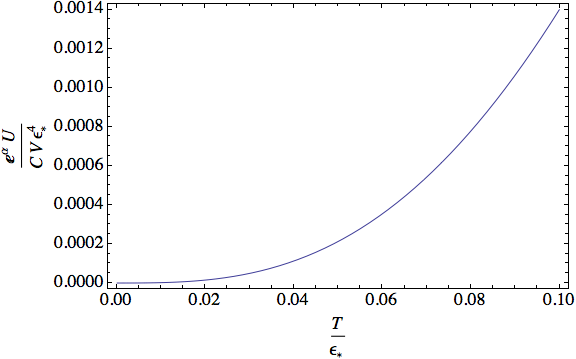}

\caption{The plot of the wealth of the system varied with economic temperature $T$. $T\in (0,0.1)$}
\label{u1}
\end{center}
\end{figure}

\begin{figure}
\begin{center}
\includegraphics[width=0.8\columnwidth]{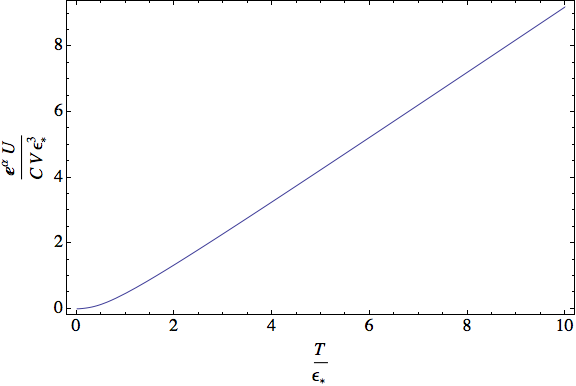}

\caption{The plot of the population of the system varied with economic temperature $T$. $T\in (0,10)$}
\label{n}
\end{center}
\end{figure}

\begin{figure}
\begin{center}
\includegraphics[width=0.8\columnwidth]{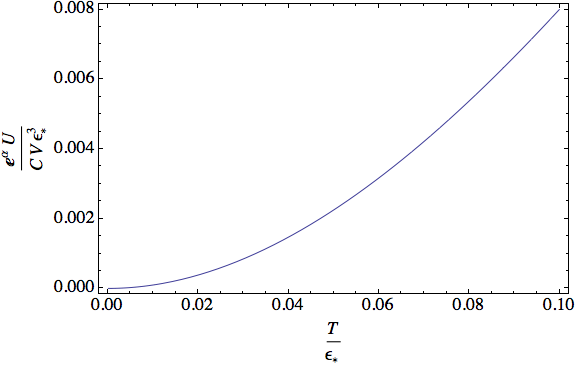}

\caption{The plot of the population of the system varied with economic temperature $T$. $T\in (0,0.1)$}
\label{n1}
\end{center}
\end{figure}

The grand partition function of this kind of system is: (could be derived from the Eq.\ref{lnz})
\begin{equation*}
ln\mathcal{Z}= \int_{0}^{\varepsilon_{*}}g(\varepsilon)e^{-\alpha -\beta \varepsilon}d\varepsilon
\end{equation*}
\begin{equation}
ln\mathcal{Z}=CV\frac{\mathrm{e}^{ - \alpha - \beta\, \varepsilon_{0}}\, \left(\beta\, \varepsilon_{*} - 2\, \mathrm{e}^{\beta\, \varepsilon_{*}} + \beta\, \varepsilon_{*}\, \mathrm{e}^{\beta\, \varepsilon_{*}} + 2\right)}{\beta^3}
\end{equation}
The wealth of the system is:
\begin{equation}
U=CV\frac{\mathrm{e}^{ -\alpha - \beta\, \varepsilon_{*}}\, \left(4\, \beta\, \varepsilon_{*} - 6\, \mathrm{e}^{\beta\, \varepsilon_{*}} + \beta^2\, \varepsilon_{*}^2 + 2\, \beta\, \varepsilon_{*}\, \mathrm{e}^{\beta\, \varepsilon_{*}} + 6\right)}{\beta^4}
\end{equation}
It is shown as Figure \ref{u} and Figure \ref{u1}.
The population of the system is:
\begin{equation}
N=CV\frac{\mathrm{e}^{ - \alpha - \beta\, \varepsilon_{*}}\, \left(\beta\, \varepsilon_{*} - 2\, \mathrm{e}^{\beta\, \varepsilon_{*}} + \beta\, \varepsilon_{*}\, \mathrm{e}^{\beta\, \varepsilon_{*}} + 2\right)}{\beta^3}
\end{equation}
The plot is shown as Figure \ref{n} and Figure \ref{n1}.

Through the comparison of Figure \ref{u}-\ref{n1}, one can get that both the wealth and population of the system increase monotonically but with different trends. The slope of increasing wealth firstly increases and then gradually decreases, while the slope of increasing population increases and get close to a constant, where the population increases linearly with the increasing of $T$.\\

The economic pressure of the system could be expressed as:
\begin{equation}
p=C\frac{\mathrm{e}^{ - \alpha - \beta\, \varepsilon_{0}}\, \left(\beta\, \varepsilon_{*} - 2\, \mathrm{e}^{\beta\, \varepsilon_{*}} + \beta\, \varepsilon_{*}\, \mathrm{e}^{\beta\, \varepsilon_{*}} + 2\right)}{\beta^4}
\end{equation}

\begin{figure}
\begin{center}
\includegraphics[width=0.8\columnwidth]{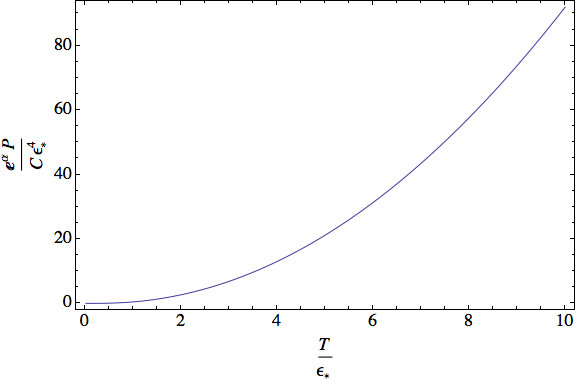}
\caption{The plot of economic pressure varied with economic temperature $T$}
\label{p}
\end{center}
\end{figure}

The trend could be seen from the Figure \ref{p} that the economic pressure which means the production efficiency increases in a more rapid way with increasing $T$. It indicates that this kind of system is a sound developing system, which demonstrates that the society which is mostly occupied by the middle class is a well-developing society. 
\subsection{The Significance of This Kind of Society}
Many economic theories\upcite{12} predict that the society  most of whose members belongs to the middle class is a stable society which is researched in this paper. From the above analysis, one can derive that the trend of the development gradually decreases from a rapid rasing trend to a stable increase because of the limitation of the maximum wealth level $\varepsilon_{*}$. Moreover, after a period of increase, the population would increase in a linear way. This is because that we assume the economic system is a similar-Boltzmann system, applying no limitation on the number of individuals on each wealth level. Through discussion, this pattern is a well-developing pattern for the society.\\

This mean that the research on this kind of system could offer a paradigm for development for the society.

\end{document}